\documentclass[journal,transmag,twoside]{IEEEtran}
\usepackage[T1]{fontenc}
\usepackage{times}
\usepackage{graphicx,color,array,cite}
\usepackage{amsfonts,amsmath,amssymb,amsbsy}
\usepackage[colorlinks=true, linkcolor=blue, citecolor=red, urlcolor=magenta]{hyperref}

\def\emph#1{\textcolor{blue}{#1}}
\begin{document}

\title{Dynamics of magnetic skyrmion clusters driven by spin-polarized current with a spatially varied polarization}

\author{\IEEEauthorblockN{%
Wenjing Jiang\IEEEauthorrefmark{1},
Jing Xia\IEEEauthorrefmark{1},
Xichao Zhang\IEEEauthorrefmark{1},
Yifan Song\IEEEauthorrefmark{1},
Chuang Ma\IEEEauthorrefmark{2}, \\
Hans Fangohr\IEEEauthorrefmark{3,4},
G. P. Zhao\IEEEauthorrefmark{5},
Xiaoxi Liu\IEEEauthorrefmark{2},
Weisheng Zhao\IEEEauthorrefmark{6},
and
Yan Zhou\IEEEauthorrefmark{1}}
\IEEEauthorblockA{\IEEEauthorrefmark{1}School of Science and Engineering, The Chinese University of Hong Kong, Shenzhen, Guangdong 518172, China}
\IEEEauthorblockA{\IEEEauthorrefmark{2}Department of Electrical and Computer Engineering, Shinshu University, Wakasato 4-17-1, Nagano 380-8553, Japan}
\IEEEauthorblockA{\IEEEauthorrefmark{3}European XFEL GmbH, Holzkoppel 4, 22869 Schenefeld, Germany}
\IEEEauthorblockA{\IEEEauthorrefmark{4}Faculty of Engineering and the Environment, University of Southampton, Southampton SO17 1BJ, United Kingdom}
\IEEEauthorblockA{\IEEEauthorrefmark{5}College of Physics and Electronic Engineering, Sichuan Normal University, Chengdu 610068, China}
\IEEEauthorblockA{\IEEEauthorrefmark{6}Fert Beijing Institute, BDBC, and School of Electronic and Information Engineering, Beihang University, Beijing 100191, China}
\thanks{W. Jiang and J. Xia contributed equally to this work. Corresponding author: Y. Zhou (email:~zhouyan@cuhk.edu.cn).}}


\IEEEtitleabstractindextext{%
\begin{abstract}
Magnetic skyrmions are promising candidates for future information
technology. Here, we present a micromagnetic study of isolated
skyrmions and skyrmion clusters in ferromagnetic nanodisks driven by
the spin-polarized current with spatially varied polarization. The
current-driven skyrmion clusters can be either dynamic steady or static,
depending on the spatially varied polarization profile. For the dynamic
steady state, the skyrmion cluster moves in a circle in the nanodisk, while for the
static state, the skyrmion cluster is static. The frequency of the circular
motion of skyrmion is also studied. Furthermore, the dependence of the
skyrmion cluster dynamics on the magnetic anisotropy and Dzyaloshinskii-Moriya interaction is
investigated. Our results may provide a pathway to realize magnetic
skyrmion cluster based devices.
\end{abstract}
}

\maketitle
\IEEEdisplaynontitleabstractindextext
\IEEEpeerreviewmaketitle

\section{Introduction}
\label{se:Introduction}

Magnetic skyrmions are topologically protected quasi-particles usually
stabilized by the Dzyaloshinskii-Moriya interaction
(DMI)~\cite{DMI_Dzyaloshinskii,DMI_Moriya,Roszler_NATURE2006,Rohart_PRB2013}. The
DMI in the bulk magnetic materials with broken inversion symmetry can
result in the formation of Bloch-type
skyrmions~\cite{Yu_NATURE2010}. The DMI originating from the interface
between ultra-thin magnetic film and a transition metal can lead to the formation of
N{\'e}el-type skyrmions~\cite{Romming_SCIENCE2013}. Because skyrmions
can have small size and reasonable stability, many potential spintronic
and electronic applications based on skyrmions have been
proposed~\cite{Fert_NNANO2013,Yan_NCOMMS2015,Xichao_SREP2015C,Upadhyaya_PRB2015,Xichao_SREP2015B,Reichhardt_NJP2015,Kang_PIEEE2016,Yuan_SREP2016,Zhang_NATCOMMUN2017,Bourianoff_AIPADV2018}. The
ability to create and observe skyrmions experimentally has further enhanced their
competitiveness~\cite{Muhlbauer_SCIENCE2009,Yu_NATURE2010,Heinze_NPHYS2011,Seki_SCIENCE2012,Romming_SCIENCE2013,Wanjun_SCIENCE2015,Wanjun_NPHYS2017}. Moreover,
skyrmions can be manipulated by a number of methods, such as
spin-polarized
current~\cite{Fert_NNANO2013,Iwasaki_NNANO2013,Woo_NMATER2016},
spin wave~\cite{Schutte_PRB2014,Xichao_NANOTECH2015}, and magnetic
field~\cite{Wang_PRB2015,Beg_SREP2015}, which diversifies the
application of skyrmions.
For the magnetic field, uniform fields in symmetric systems cannot move skyrmions
but one needs to break the in-plane symmetry, such as applying an additional
in-plane field~\cite{Wang_PRB2015} or a field gradient~\cite{Buttner_NPHYS2015,Liang_ARXIV2017}.

In this \textit{Letter}, we show that the isolated skyrmion and
skyrmion cluster~\cite{Zhao_PNAS2016,Rozsa_PRL2017,Pepper_ARXIV2018} in a ferromagnetic nanodisk with
interface-induced DMI can be manipulated by the spin-polarized current
with a spatially varied polarization, e.g., a vortex-like
polarization. Such a vortex-like polarization is assumed to be
realized by a vortex polarizer, which has been used to improve the
performance of nano-oscillators in previous
literature~\cite{Carpentieri_JAP2011,Kim_NJP2016}. The skyrmions
driven by the spin-polarized current with a vortex-like polarization
can reach either a dynamic steady state or a static state, depending on the
profile of the spatially varied polarization. With applying the vertical
spin current, some of the skyrmions in the skyrmion cluster are destroyed.
For the dynamic steady state, the remaining skyrmions move in a circle in the final state,
while for the static state, the remaining skyrmions are static.
Hence, using the spin-polarized current with a vortex-like polarization,
it is possible to rotate, assemble, compress and delete skyrmions in the nanodisk in
a quantitative manner.

\begin{figure}[t]
\centerline{\includegraphics[width=0.35\textwidth]{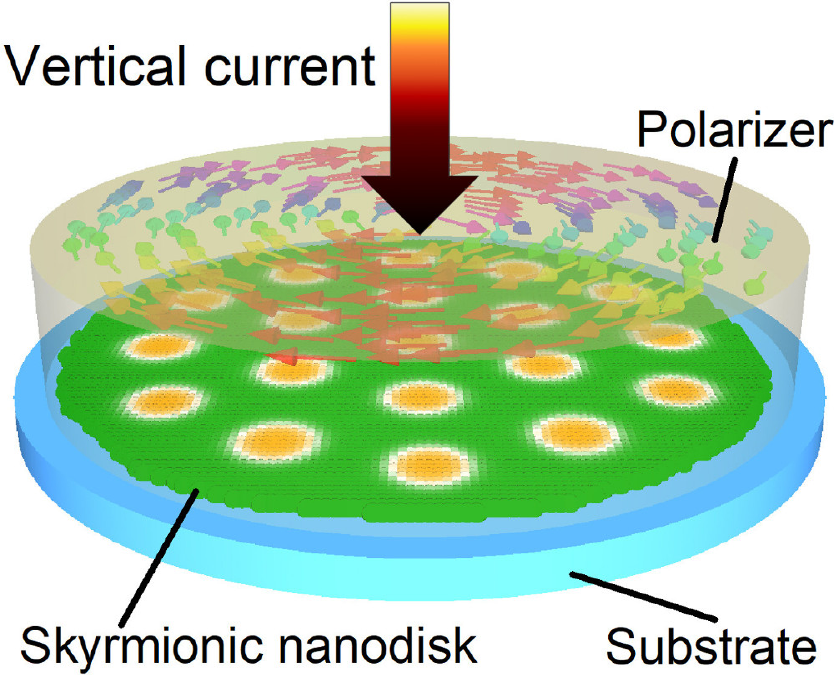}}
\caption{Illustration of the simulation model. The vortex spin
  polarizer is in contact with an ultra-thin ferromagnetic
  nanodisk. The interface between the nanodisk and the heavy-metal
  substrate provides the DMI. The skyrmions in the nanodisk are driven
  by the spin current vertically injected through the polarizer.}
\label{FIG1}
\end{figure}

\begin{figure}[t]
\centerline{\includegraphics[width=0.50\textwidth]{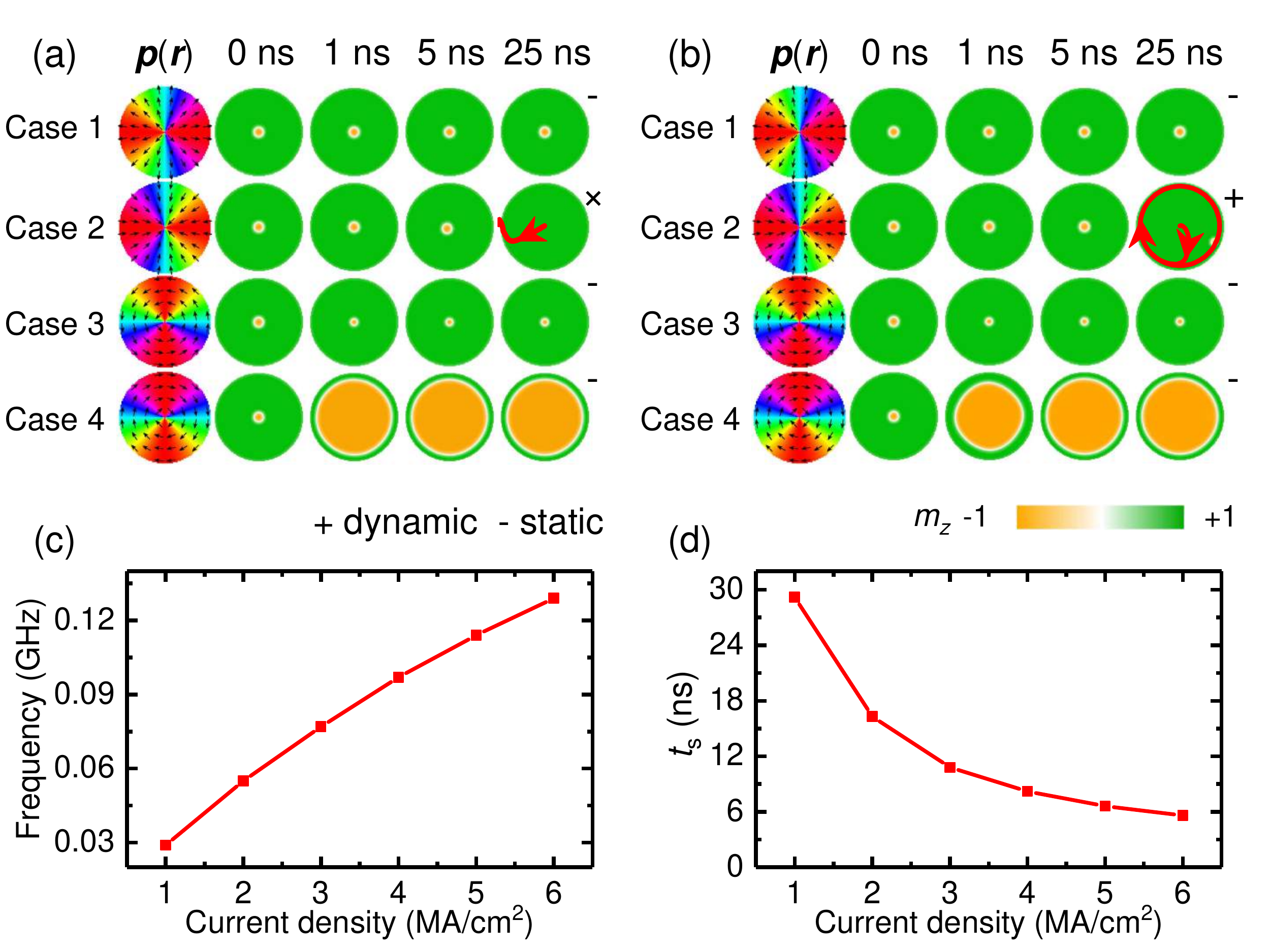}}
\caption{Motion of an isolated skyrmion driven by the vertically
  injected spin current with spatially varied polarization in a
  nanodisk with (a) $\alpha=0.03$ and (b) $\alpha=0.30$. (c) The
  oscillation frequency of skyrmion and (d) starting oscillation time
  for the case 2 of (b). The PMA constant $K = 0.8$ MJ/m$^3$, DMI
  constant $D=3.5$ mJ/m$^2$, and current density $j = 5$
  MA/cm$^2$. The symbol \textbf{-}, \textbf{+}, and \textbf{$\times$}
  indicate the skyrmion is static, dynamic, and broken in the final
  stable state, respectively. The trajectories of skyrmion are given
  in the case 2 of (a) and (b), where the arrow indicates the motion
  direction.}
\label{FIG2}
\end{figure}

\section{Methods}
\label{se:Methods}

The simulation model, as shown in Fig.~\ref{FIG1}, is an ultra-thin
ferromagnetic nanodisk with the thickness of $1$ nm and the diameter
of $300$ nm on a heavy-metal substrate inducing DMI. The micromagnetic
simulations are performed with the Object Oriented MicroMagnetic
Framework (OOMMF) simulator~\cite{OOMMF}. The dynamics of
magnetization is described by the Landau-Lifshitz-Gilbert (LLG)
equation including a Slonczewski-like torque, written as
\begin{equation}
\frac{d\boldsymbol{m}}{dt}=-\gamma_{0}\boldsymbol{m}\times\boldsymbol{h}_{\rm{eff}}+\alpha(\boldsymbol{m}\times\frac{d\boldsymbol{m}}{dt})-u\boldsymbol{m}\times(\boldsymbol{m}\times\boldsymbol{p}),
\label{eq:LLG}
\end{equation}
where $\boldsymbol{m}$ is the reduced magnetization
$\boldsymbol{M}/{M_\text{S}}$, $M_\text{S}$ is the saturation
magnetization, $\gamma_{0}$ is the gyromagnetic ratio and $\alpha$ is
the damping coefficient, $\boldsymbol{h}_{\text{eff}}$ is the
effective field including the contributions of Heisenberg exchange,
DMI, perpendicular magnetic anisotropy (PMA) and demagnetization
field. The variable $u$ is defined as
$(\gamma_{0}\hbar jP)/(2ae\mu_{0}M_{\text{S}})$, where $\hbar$ is the
reduced Plank constant, $j$ is the current density, $P=1.0$ is the
assumed full spin polarization rate, $a$ is the nanodisk thickness,
$e$ is the electron charge, $\mu_{0}$ is the vacuum permeability
constant, and $\boldsymbol{p}$ is the spin polarization. We consider four
cases of the spatially varied polarization profile $\boldsymbol{p}(\boldsymbol{r})$, 
which could be realized by a vortex polarizer, as shown in 
Fig.~\ref{FIG2} (a) in the left-most column.
While the field-like torque can occur in relevant systems, it is not included in this study because we have checked and found that the effect of field-like torque on the skyrmion dynamics is small.

The model is discretized into cuboidal volume elements with the size
of $3$ nm $\times$ $3$ nm $\times$ $1$ nm. The default DMI constant 
is $3.5$ mJ/m$^2$. For the phase diagrams, DMI constant 
varies from $2.5$ mJ/m$^2$ to $4.5$ mJ/m$^2$ with a step of $0.5$ mJ/m$^2$. 
The interfacial DMI strength can be controlled by varying the thickness 
of the underlying heavy metal layer~\cite{Tacchi_PRL2017}. The saturation 
magnetization is $M_{\text{S}} = 580$ kA/m. The exchange constant is
$A = 15$ pJ/m. The PMA constant $K$ varies from $0.6$ MJ/m$^3$ to
$1.0$ MJ/m$^3$. The Gilbert damping coefficient $\alpha$ ranges $0.01$
to $1.0$. The spin-polarized current injection area is a circle with a
diameter of $300$ nm upon the nanodisk. The current density $j$ is in
the range of $1 \sim 20$ MA/cm$^2$. The Oersted field is not included for all the simulations.
Initially, an isolated skyrmion or skyrmion cluster has been created in the nanodisk and relaxed into stable state. The skyrmion cluster is composed of $19$ skyrmions in
our simulations.

\begin{figure}[t]
\centerline{\includegraphics[width=0.50\textwidth]{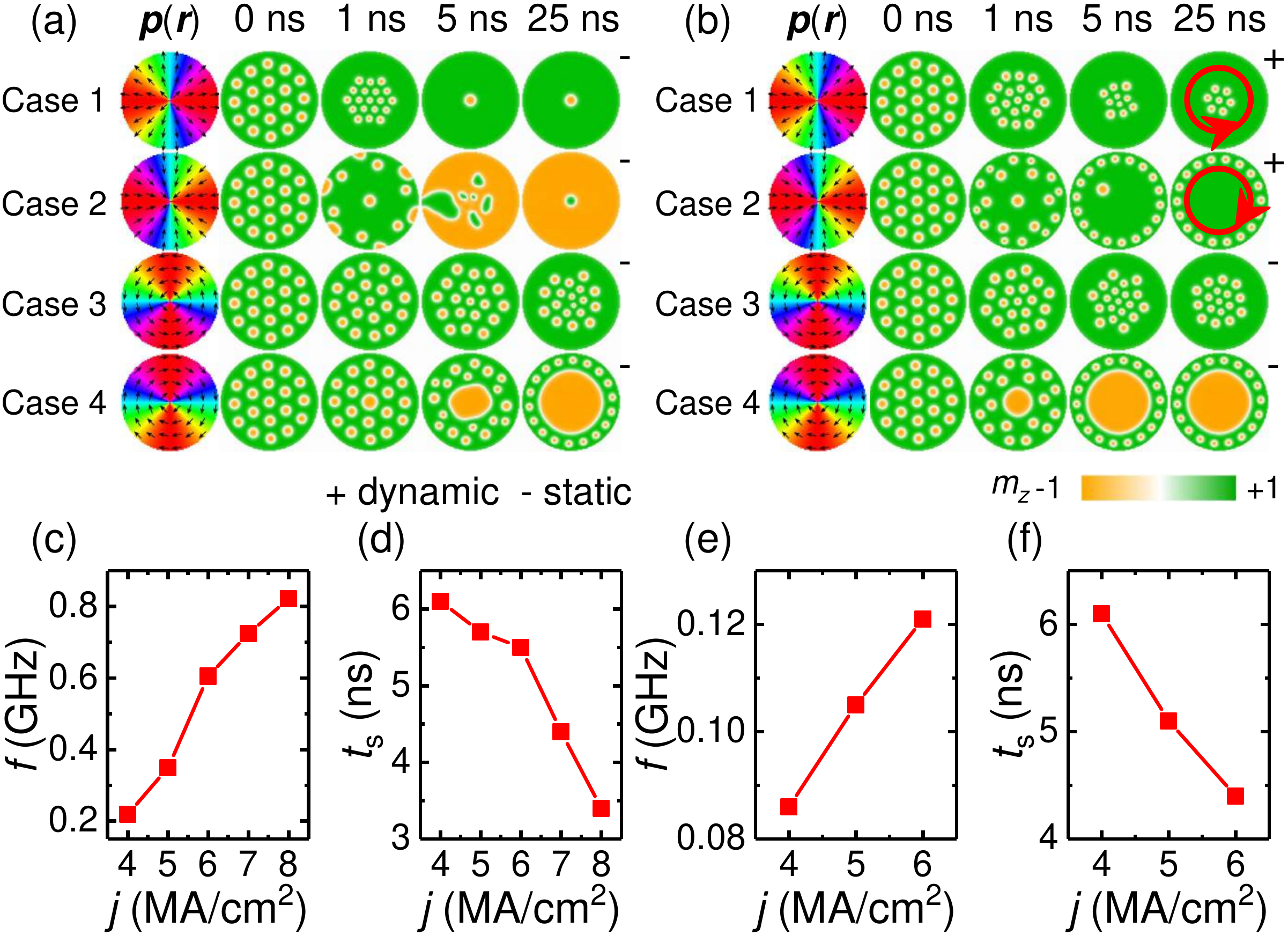}}
\caption{Motion of a skyrmion cluster driven by the vertically
  injected spin current with spatially varied polarization in a
  nanodisk with (a) $\alpha=0.03$ and (b) $\alpha=0.30$. (c) The
  oscillation frequency of skyrmion and (d) starting oscillation time
  for the case 1 of (b). (e) The oscillation frequency of skyrmion and
  (f) starting oscillation time for the case 2 of (b). The PMA
  constant $K = 0.8$ MJ/m$^3$, DMI constant $D=3.5$ mJ/m$^2$, and
  current density $j = 5$ MA/cm$^2$. The symbol \textbf{-} and
  \textbf{+} indicate the final states are static and dynamic,
  respectively. The red arrows represent directions of skyrmion
  cluster rotation.}
\label{FIG3}
\end{figure}

\section{Results and Discussion}
\label{se:Results}

\subsection{Manipulating the isolated skyrmion}
\label{se:isolatedSk}

We first study the manipulation of an isolated skyrmion driven by the
spin current with the spatially varied polarization. Initially, an
isolated skyrmion is placed at the nanodisk center and relaxed. The
results for $\alpha=0.03$ and $0.30$ are shown in Fig.~\ref{FIG2}(a)
and (b), respectively. The case 1 and 2 are corresponding to the
outward-pointing and inward-pointing radial polarization profiles,
respectively. The case 3 and 4 corresponds to the counterclockwise and
clockwise vortex-like polarization profiles, respectively. For case 1,
the skyrmion doesn't move when the spin current is injected. For case
2, the skyrmion stays at its initial position for $t<4$ ns. It can be
seen from the second row of Fig.~\ref{FIG2}(a), the skyrmion shifts
from the center of the nanodisk slightly when $t=5$ ns. Finally, the
skyrmion is destroyed by touching the nanodisk edge. The trajectory of
the skyrmion is indicated by the red curve. For case 3, when the spin
current is applied, the radius of skyrmion shrinks from $15.9$ nm to
$11.2$ nm. For case 4, the skyrmion size expands and the size is
comparable to the size of nanodisk finally. In Fig.~\ref{FIG2}(b), for
$\alpha=0.30$, similar results can be found for different polarization
profiles but for case 2, where the skyrmion steadily moves in a circle
along the nanodisk edge at last. Figure~\ref{FIG2}(c) and (d) shows
the frequency and starting time of the steady circular motion of skyrmions at
various current density. Increasing current density results in faster
motion of skyrmion, leading to the increase of frequency. In Fig.~\ref{FIG2}(d), the starting time when the amplitude of the displacement is $2$ nm decreases with increasing current density.

The behaviors of the skyrmion can be explained via Thiele equation. The Thiele motion equation is derived from LLG equation, which is expressed as
\begin{equation}
-G\hat{z}\times\boldsymbol{v}+ \alpha\boldsymbol{D}\cdot\boldsymbol{v}-u\boldsymbol{I}\cdot \boldsymbol{p}=0,
\label{TME}
\end{equation}
where $D_{xx}=D_{yy}=D$ and $0$ otherwise, $I_{xy}=-I_{yx}=I$ and $0$ otherwise. The constant $G, D$ and $I$ are determined by the profile of skyrmion. For the skyrmion in this paper, $G$ is negative, $D$ and $I$ are positive. Then, the velocity can be obtained and given by
\begin{equation}
v_x=uI\frac{G p_x-\alpha D p_y}{G^2+\alpha^2D^2}, v_y=uI\frac{\alpha D p_x+G p_y}{G^2+\alpha^2D^2}.
\label{v}
\end{equation}
When the damping is small, then the velocity can be expressed as
\begin{equation}
v_x=uI p_x/G, v_y=uI p_y/G.
\label{v-smalldamping}
\end{equation}
For case 1, the polarization profile is radically outward. $G$ is negative,
resulting in that the $\boldsymbol{v}$ is radically inward. Initially,
a skyrmion is placed in the center of the disk, then the skyrmion keeps its
location when the spin current is applied. For case 2, the polarization
profile is radically inward, leading to the radically outward motion of skyrmion.
From the case 2 in Fig.~\ref{FIG2}(a), we can see that the skyrmion moves radically outward
and is destroyed by touching the edge. When the damping is increased to $0.3$,
the damping term in Eq.~\ref{v} cannot be neglected, resulting in a spiral inward
and outward velocity profile for case 1 and case 2 respectively. Hence,
the skyrmion for case 2 moves in a curved trajectory before reaching the edge,
which is different from the case of $\alpha=0.03$ where skyrmion moves in a
straight line before reaching the edge. Due to the spiral outward driving force
and the edge repulsion, the skyrmion reaches a steady circular motion.

For case 3 and 4, only static states are found. The direction of
effective field resulted by Slonczewski-like torque is $\boldsymbol{m}\times\boldsymbol{p}$.
For case 3, the effective field exerting on the domain wall of skyrmion is pointed up, which leads to the shrinking of skyrmion. For case 4, the effective field pointing down results in the expansion of skyrmion.

\subsection{Manipulating the skyrmion cluster}
\label{se:SkCluster}

We continue to study the manipulation of a skyrmion cluster including
$19$ skyrmions driven by the spin current with the spatially varied
polarization. The behaviors and the final states of the skyrmion
cluster are given in Fig.~\ref{FIG3}. In Fig.~\ref{FIG3}(a), $\alpha$
is $0.03$. For case 1, the radially inward driving force and the repulsion
between skyrmions result in the destruction of skyrmions and
only one single skyrmion centered at the nanodisk survives. In case 2,
the skyrmions are rapidly expelled out of the nanodisk except for the
center one. Then, the center one is expanded and transformed to become
multiple domains. Eventually, a skyrmion with a core pointing the $+z$
direction is formed at the center of the nanodisk. It remains as a static
state. For case 3, the skyrmion cluster is compressed toward the
nanodisk center. Some of skyrmions in the nanodisk center shrink and
are destroyed due to the strong repulsion between the skyrmions. For
this case, there are $17$ skyrmions left at $t=25$ ns. In the final
static state at $t=50$ ns, we find that $13$ skyrmions remain.
In case 4, the skyrmion in the center expands and other skyrmions
move radially outward. At last, the skyrmion in the center is
expanded significantly and others are driven toward the boundary and
distributed uniformly along the nanodisk edge.

The large damping coefficient ($\alpha=0.3$) results in a dynamic
steady state for case 1 and 2, as shown in Fig.~\ref{FIG3}(b). For
case 1, the skyrmions move toward the nanodisk center and
some of them are destroyed due to the spiral inward driving force
and repulsion between skyrmions. We find $7$ skyrmions are left with
$6$ skyrmions moving around the center one. In case 2, skyrmions are
driven towards the boundary. Because the space is limited, one skyrmion
is expelled out the nanodisk. The other $18$ skyrmions are distributed
uniformly along the edge and driven into clockwise circular motion,
settling into a steady dynamic state. For
cases 3 and 4, the results are similar to these in
Fig.~\ref{FIG3}(a). The static states are obtained. For case 1 and 2
in Fig.~\ref{FIG3}(b) where the dynamic stable states are obtained in
the end, the frequency and starting time of the circular motion of skyrmions as
functions of the current density are shown in Fig.~\ref{FIG3}(e)-(f).
For the skyrmion cluster in case 1 and case 2, some skyrmions are
destroyed at the beginning and then the remaining skyrmions move
in a circle. The starting time is the time when the steady circular
motion of remaining skyrmions occurs.

\begin{figure}[t]
\centerline{\includegraphics[width=0.50\textwidth]{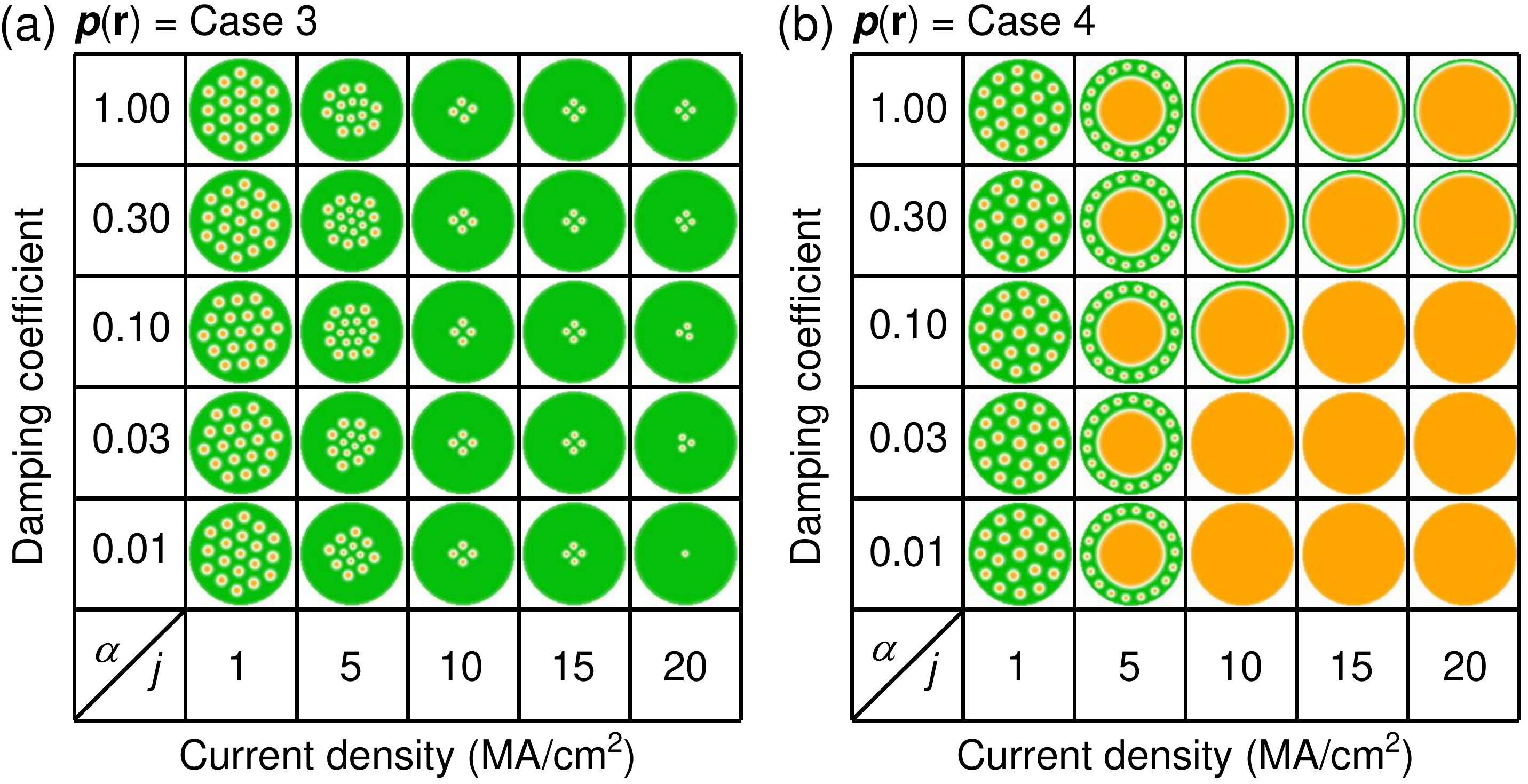}}
\caption{Final states of skyrmion clusters driven by the spin
  current polarized with the polarization profiles of (a) case 3, (b)
  case 4 at different current densities and damping coefficients. The
  PMA constant $K = 0.8$ MJ/m$^3$ and DMI constant $D=3.5$ mJ/m$^2$.}
\label{FIG4}
\end{figure}

The final stable states of skyrmion clusters consisting of $19$
skyrmions driven by the spin current with vortex-like polarizations
(case 3 and case 4) for various current density $j$ and damping
coefficient $\alpha$ are shown in Fig.~\ref{FIG4}. For case 3, it can
be seen that the number of remaining skyrmions decreases when the
applied current density is increased. For example, for $\alpha=0.01$,
$19$ skyrmions are displaced slightly toward nanodisk center when the
spin current density $j = 1$ MA/cm$^2$, while only one skyrmion exists
when $j = 20$ MA/cm$^2$. More skyrmions are left in the final stable
state when damping constant increases. For example, when $j = 20$
MA/cm$^2$, there is only one skyrmion in the final stable state for
$\alpha=0.01$ and $4$ skyrmions for $\alpha=1.0$. It is noteworthy
that the distributions of skyrmions in the final static state are
symmetric. For example, when $j = 1$ MA/cm$^2$, the skyrmions are
arranged as hexagon lattice for $\alpha=0.01\sim 1.0$.

\begin{figure}[t]
\centerline{\includegraphics[width=0.50\textwidth]{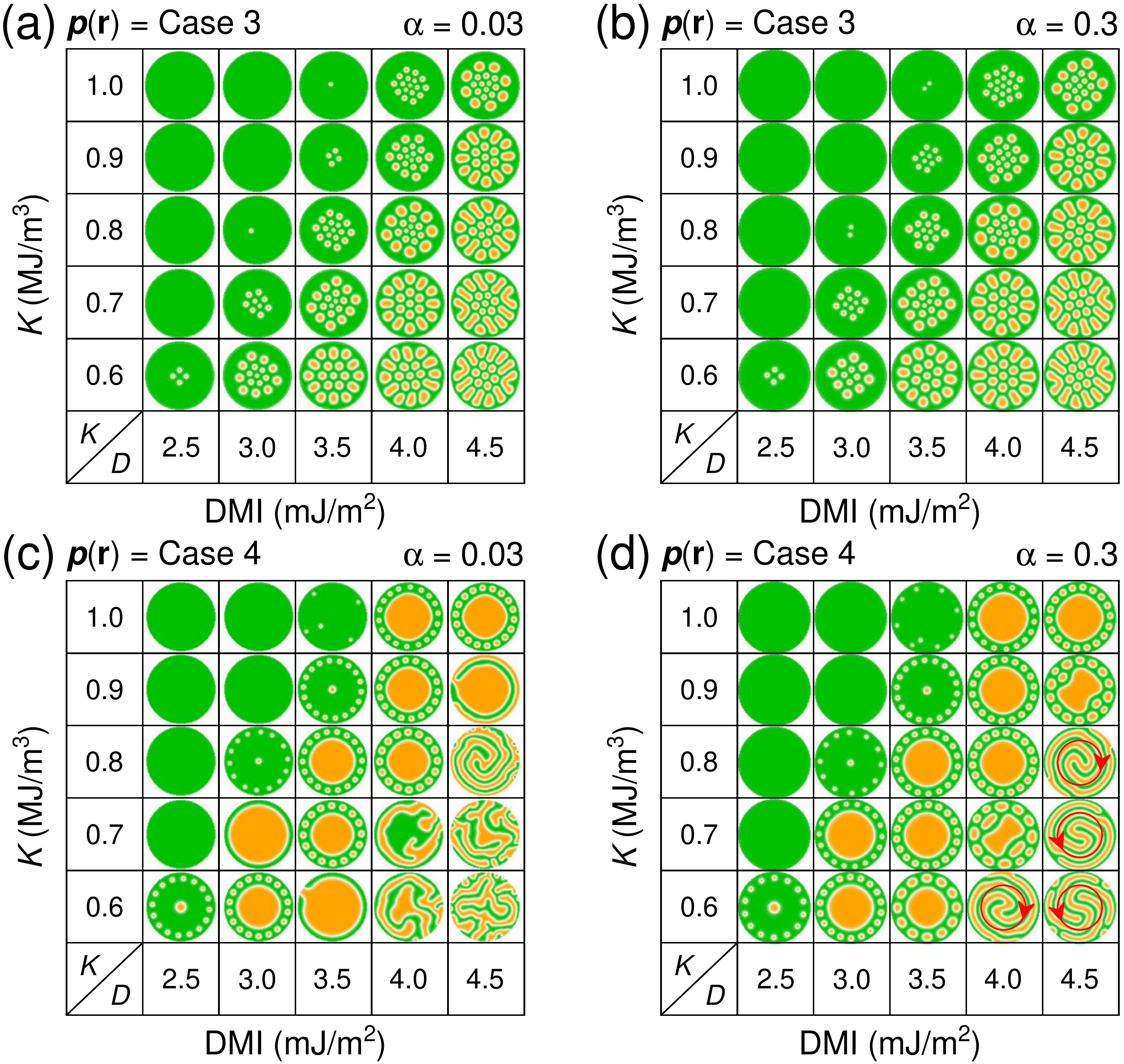}}
\caption{Final states of skyrmion clusters driven by the spin current as
  functions of DMI and PMA for different damping coefficient and
  polarization profile. (a) $\alpha=0.03$ for case 3. (b)
  $\alpha=0.30$ for case 3. (c) $\alpha=0.03$ for case 4. (d)
  $\alpha=0.30$ for case 4. The current density $j = 5$ MA/cm$^2$. The
  arrows represent the rotation direction of the magnetization
  structures.}
\label{FIG5}
\end{figure}

For case 4, the static final stable states for various current density
$j$ and damping constant $\alpha$ are shown in Fig.~\ref{FIG4}(b). The
injection of the spin current with the polarization profile of case 4
results in the expansion of the centered skyrmion. When the applied current density is small,
$j = 1$ MA/cm$^2$ and $\alpha=0.01$, the expansion of the skyrmion in
the center and the displacement of the other skyrmions are small. When
the current density is increased to $5$ MA/cm$^2$, the expansion of
the skyrmion in the center is obvious and the other skyrmions are
pushed to the nanodisk edge. When the current density is further
increased to $10$ MA/cm$^2$, the domain with magnetization pointing
down is expanded over the nanodisk. Therefore, a uniform
ferromagnetic state is formed in the nanodisk. When $\alpha=0.10$, for
$j=10$ MA/cm$^2$, the driving force of the spin current and the
expansion of the skyrmion at the nanodisk center erase all the other
skyrmions. The expansion of the skyrmion in the center is stopped by
the interaction from the edge. At last, an expanded skyrmion is
obtained. When $j$ is increased to $15$ MA/cm$^2$, the large driving
force leads to a significant expansion of the domain and a uniform
ferromagnetic state is obtained.

The phase diagram as functions of the DMI constant $D$ and the PMA
constant $K$ at a fixed damping coefficient of $\alpha=0.03$ or
$\alpha=0.3$ for case 3 and case 4 are shown in Fig.~\ref{FIG5}. All
phase diagrams presented in Fig.~\ref{FIG5} exhibit the single domain
for the large $K$ and small $D$ since the critical DMI constant
$D_c=4\sqrt{AK}/\pi$~\cite{Rohart_PRB2013}. When $D$ is much smaller
than $D_c$, the skyrmion cannot be stabilized by the DMI in the
nanodisk. For case 3, as shown in Fig.~\ref{FIG5}(a) and (b), the
final stable states are static for $\alpha=0.03$ and $0.3$. The size
of skyrmions increases as the DMI constant increases.

For case 4 and $\alpha=0.03$, when $K=0.6$ MJ/m$^3$, the expansion of
the skyrmion in the center is not significant for $D=2.5$ mJ/m$^2$. For
larger $D$, the expansion is remarkable, as shown in
Fig.~\ref{FIG5}(c). When $D$ increases to $4.0$ mJ/m$^2$, the final
state contains multiple domains and all skyrmions are destroyed. For
case 4 and $\alpha=0.30$, four dynamic stable states of domain walls are found from
the phase diagram, as indicated with red directed loops in
Fig.~\ref{FIG5}(d), where the multiple domain walls rotate clockwise
or counterclockwise driven by the spin current. Other results are
static in the final stable states similar to those results in the
Fig.~\ref{FIG5}(c).

\section{Conclusion}
\label{se:Conclusion}

We have studied the motion of magnetic skyrmions driven by
spin-polarized currents with different spatially varying
polarizations. The skyrmions move radically inward (case 1) or outward (case 2)
when the damping is small, depending on the polarization profile.
Due to the repulsions between skyrmions and edge repulsions, some of skyrmions
may be destroyed and the remaining skyrmions are static in the final state.
When the damping is large, the skyrmions move spirally inward (case 1)
or outward (case 2) and the remaining skyrmions move in a circle in the final.
The spin-polarized currents with the polarization profile of case 3 and case 4
lead to the shrink and expansion of the skyrmion cluster respectively,
resulting in the deleting of skyrmions. In the final, the remaining skyrmions
are static. Our results could provide a way to manipulate skyrmion clusters in
future skyrmion-based devices.

\section*{Acknowledgment}

X.Z. was supported by the JSPS RONPAKU (Dissertation Ph.D.) Program.
G.P.Z. was supported by the National Natural Science Foundation of China (Grant Nos. 51771127, 11074179 and 10747007), and the Construction Plan for Scientific Research Innovation Teams of Universities in Sichuan (Grant No. 12TD008).
W.Z. acknowledges the support by the projects from National Natural Science Foundation of China (Grant Nos. 61501013, 61571023 and 61627813), the International Collaboration Project from the Ministry of Science and Technology in China (Grant No. 2015DFE12880), and the Program of Introducing Talents of Discipline to Universities in China (Grant No. B16001).
Y.Z. acknowledges the support by the President's Fund of CUHKSZ, the National Natural Science Foundation of China (Grant No. 11574137), and Shenzhen Fundamental Research Fund (Grant Nos. JCYJ20160331164412545 and JCYJ20170410171958839).




\end{document}